\documentclass[12pt]{article}
\usepackage{float}
\usepackage{makecell}
\usepackage{subfig}
\usepackage[hyphens]{url}
\usepackage{multirow}
\usepackage{hhline}
\usepackage{subfig}
\usepackage[latin2]{inputenc}
\usepackage{changepage}
\usepackage{amsmath}
\usepackage[pdftex]{graphicx}
\usepackage{amsfonts}
\usepackage{amssymb}
\usepackage[left=1.5in]{geometry}
\usepackage{geometry}
\usepackage{wrapfig}
\usepackage{setspace}
\usepackage{epsfig}
\usepackage{color}
\usepackage{tikz}
\usepackage{cancel}
\usepackage{float}
\usepackage{cite}
\usepackage{url}
\usepackage[hidelinks]{hyperref}
\usepackage{authblk}
\usepackage{lineno}
\title{ Effects of Social Cues on Biosecurity Compliance in Livestock Facilities: Evidence from Experimental Simulations}
\author[1,2]{Luke Trinity}
\author[1,3,4]{Scott C. Merrill}
\author[1,2]{Eric Clark}
\author[1,4,5]{Christopher J. Koliba}
\author[1,4,5]{Asim Zia}
\author[1,3]{Gabriela Bucini}
\author[1,6]{Julia M. Smith}
\affil[1]{{\tiny SEGS Lab, The University of Vermont, Burlington, VT, 05401}}
\affil[2]{{\tiny Complex Systems Center, The University of Vermont, Burlington, VT, 05401}}
\affil[3]{{\tiny Department of Plant and Soil Science, The University of Vermont, Burlington, VT, 05401}}
\affil[4]{{\tiny Gund Fellow, Gund Institute for Environment, University of Vermont, Burlington, VT, 05401}}
\affil[5]{{\tiny Department of Community Development and Applied Economics, University of Vermont, Burlington, VT, 05401}}
\affil[6]{{\tiny Department of Animal and Veterinary Sciences, University of Vermont, Burlington, VT, 05401}}
\begin{document}

\maketitle

\newpage
\begin{abstract}
\vspace{10mm}
Disease outbreaks in U.S. animal livestock industries have economic impacts measured in hundreds of millions of dollars per year. Biosecurity, or procedures intended to protect animals against disease, is known to be effective at reducing infection risk at facilities. Yet to the detriment of animal health, humans do not always follow biosecurity protocols. Human behavioral factors have been shown to influence willingness to follow biosecurity protocols. Here we show how social cues may affect cooperation with a biosecurity practice. Participants were immersed in a simulated swine production facility through a graphical user interface and prompted to make a decision that addressed their willingness to comply with a biosecurity practice. We tested the effect of varying three experimental variables: (1) the risk of acquiring an infection, (2) the delivery method of the infection risk information (numerical versus graphical), and (3) behavior of an automated coworker in the facility. We provide evidence that participants changed their behavior when they observed a simulated worker making a choice to follow or not follow a biosecurity protocol, even though the simulated worker had no economic effect on the participants' payouts. These results advance the understanding of human behavioral effects on biosecurity protocol decisions; demonstrating that social cues need to be considered by livestock facility managers when developing policies to make agricultural systems more disease resilient.

\end{abstract}

\section{Introduction}

Endemic and emergent diseases remain a constant threat to the animal and economic welfare of the livestock industry. A national survey of U.S. hog producers found that from 2014-2017, 54.3\% reported having suffered an outbreak of Porcine reproductive and respiratory syndrome virus (PRRSv), and 43.9\% reported undergoing an outbreak of Porcine epidemic diarrhea virus (PEDv). \cite{pudenz2017biosecurity} Annual economic losses due to PRRSv and PEDv are estimated at \$580.62 million \cite{porkboard2017}, and upwards of \$900 million \cite{PaarlbergPhilip2014UEEW}, respectively. Animal health issues become even more pronounced considering the current threat of African swine fever, a highly virulent disease that can cause up to 100\% fatality in pigs. \cite{sanchez2018african} Furthermore, analysis of consumer preferences indicates that animal welfare is an important attribute to some consumers of livestock products. \cite{cummins2016understanding} During disease outbreaks, public concerns related to food safety can cause reductions in pork consumption, which carries economic ramifications for the swine industry. Implementation of biosecurity best management organizational policies is critical to effectively prevent or control outbreaks of existing and emerging virulent diseases. \cite{ritter2017invited}

Human behavioral factors that can influence biosecurity implementation have been identified as crucial to mitigating the risk of outbreaks; although limited knowledge exists on the actual relationship between these factors and behavior. \cite{hidano2018, MankadAditi2016Piob} Analysis of human-behavior can be undertaken in contexts of varying scope: strategic, tactical, and operational. \cite{schmidt2000strategic} Taking a broad view, big picture strategic decisions are guided by long-term objectives. Biosecurity issues at the strategic scale are often related to network interactions of facilities and service providers across production chains. The emergent behavior of networks, an important consideration of strategic policy makers, is intrinsically related to more localized tactical decisions that are made within a specified region. At the tactical level, farm managers decide whether or not to invest in and implement preventive biosecurity protocols. From a narrower, more localized perspective, operational level biosecurity can be viewed as a continuous series of decisions made by production workers indicating their willingness to follow or comply with biosecurity protocols (hereafter referred to as compliance with biosecurity). \cite{LoorbachDerk2010TMfS} The operational, tactical, and strategic levels of biosecurity are interconnected, for example, tactical decision-making influences operational level protocols. And while it is known that operational level compliance with biosecurity will impact the implementation and efficacy of tactical and strategic biosecurity decisions, limited feedback describing operational level behavior is currently available to inform tactical and strategic level decision-making. 

Compliance with biosecurity, such as consistently following sanitation protocols before entering a production facility, has been reported to significantly reduce disease. \cite{AndresVictorM.2015BMtC, NoremarkMaria2016SFOa, beloeil2007risk} Unfortunately, poor compliance with biosecurity is an endemic problem in many animal production systems. A detailed questionnaire of 60 Swedish farrow-to-finish herds found that particular biosecurity measures were applied for visitors in more than half of farms; but those same measures were carried out by farmers and staff themselves on only 32\% of farms. \cite{BackhansAnnette2015Blah} Hidden cameras at Quebec poultry farms documented 44 different biosecurity lapses made by workers and visitors over a four-week period. \cite{RacicotManon2011Do4b} Workers may be exposed to a variety of information about the consequences of a facility infection; but they balance the cost of infection with complacency and a tendency to become lax in day-to-day activities. This relaxation of biosecurity effort has been examined from a temporal perspective, with evidence suggesting that people view the likelihood and impact of an event to be reduced as time since the event increases. \cite{CarusoEugeneM2008AWiT, trope2003temporal, YiRichard2006DoPO} Referred to as temporally-based psychological distancing, the farther in the past an event occurs, the less likely and impactful such an event is perceived to be. Pressure to complete work efficiently with time constraints has created scenarios where workers find it unrealistically challenging to complete their job while complying with biosecurity standards. \cite{millman2017} A better understanding of the complex human decision-making process that influences workers' willingness to comply with biosecurity protocols requires innovative research approaches and data collection techniques that can provide novel feedback for tactical and strategic level decision makers. 

Studies of human behavioral strategies have applied serious games developed for data collection or education as early as 1962 for designing robot mining simulations. \cite{toda1962the} Experimental economic games are a particular type of serious game in which participants are incentivized with monetary payouts. Performance-based incentives are known to increase engagement and salience in decision-making. \cite{ SmithVernonL.1976EEIV, CamererColin1999TEoF, CheongLisa2016Etio} Indeed, Holt \& Laury found in their multiple price lottery experiment that risk aversion increased when human subjects were faced with real financial incentives. \cite{holt2002risk} Computer-based simulations provide a unique opportunity to study the mechanics of decision-making in a controlled environment. \cite{porter1995computer} Experiments utilizing computer software have had success amongst adolescents, increasing empathy \cite{boduszek2019prosocial}, as well as awareness of substance use \cite{montanaro2015using}. Within the domain of animal biosecurity, serious games have explored the effects of information awareness and audience on tactical investments \cite{Merrill2019PloS, clark2019using}, as well as the effects of message delivery method on operational compliance \cite{Merrill2019FVS}. 

Recent research conducted by Merrill et al. \cite{Merrill2019FVS} utilized an experimental simulation of a livestock production facility to examine factors that may influence perception of disease risk, thus affecting biosecurity compliance. Their experimental treatment variables included: information regarding disease infection risk, uncertainty associated with disease infection risk information, and the message delivery method used to communicate disease infection risk. These factors have been identified as important within the farmer biosecurity decision-making process. \cite{hidano2018, Ellis-IversenJohanne2010Pcam} Research has suggested that farmers in the United States exhibit risk tolerance \cite{roe2015}, but they may be more likely to implement or comply with biosecurity as their perceived risk of infection increases \cite{ ritter2017invited, AjzenIcek1991Ttop, Merrill2019PloS}. Indeed, Merrill et al. \cite{Merrill2019FVS} found that as the actual infection risk within the experimental simulation increased, so did compliance with a biosecurity practice. The certainty of infection risk information is also expected to affect biosecurity implementation and compliance \cite{ritter2017invited}; although the effect may change depending on the domain of interest, i.e. tactical versus operational. Merrill et al. \cite{Merrill2019PloS} found that in a \textit{tactical} experimental simulation of disease in a swine production region, an increase in disease risk certainty was associated with increased biosecurity implementation. However, in an \textit{operational} experimental simulation of a single livestock production facility, an increase in certainty was associated with decreased biosecurity compliance. \cite{ Merrill2019FVS} The opposite responses to increases in disease risk certainty at the tactical and operational levels highlights the complexity of human behavior, and the need to provide operational feedback to tactical and strategic decision makers which is currently unavailable.

Another factor that is expected to impact compliance with biosecurity protocols is the message delivery method when communicating disease risk level. \cite{LipkusI.M.1999Tvco} The method of delivering a message is important because humans exhibit affective reasoning: feelings and initial reaction to stimuli guide decision-making. \cite{SlovicPaul2004RaAa, SlovicPaul2005ARaD} Additionally, messaging affects the balance of experiential and analytical reasoning in the processing of statistical information. \cite{MarxSabineM.2007Camp} Humans rely on a limited number of mental heuristics, or ways to reduce problem complexity, when presented with decisions under uncertainty. \cite{TverskyA1974JuUH} Visual or graphical communication of risk can be advantageous in attracting and holding attention. \cite{LipkusI.M.1999Tvco} Indeed, Merrill et al. \cite{Merrill2019FVS} found that risk information delivered in a graphical format was more effective in increasing biosecurity compliance, with respect to messages delivered in a numeric or linguistic format. 

Contextual and situational factors also determine how probability is understood. \cite{patt2003using, visschers2009probability } Generally, humans underestimate risks that occur frequently \cite{scholderer2019social}, and discount the risk of an event if the probability is low \cite{KahnemanDaniel1979PTAA }. However, the exact definition of \textit{low} is malleable and subjective, as it can change depending on context. Humans utilize mental shortcuts when making decisions with limited time. One such common heuristic is referred to as anchoring and adjustment. \cite{TverskyA1974JuUH} The following example illustrates how the formulation of a problem can profoundly impact the final answer. When two groups of high school students were asked to estimate a numerical expression within 5 seconds, they used extrapolation and adjustment to formulate their answer under time pressure. The first group, when asked to estimate $1 \times 2 \times 3 \times 4 \times 5 \times 6 \times 7 \times 8$ yielded a median estimate of 512; while the second group, presented with the expression $8 \times 7 \times 6 \times 5 \times 4 \times 3 \times 2 \times 1$ yielded a median estimate of 2,250. \cite{TverskyA1974JuUH} Although the correct answer is 40,320 in both cases, changing the initial point of reference completely alters the result of the off-hand mental calculation. 

In this study we expand upon the work of Merrill et al. \cite{Merrill2019FVS} by introducing a novel variable to treat: information about social cues in a production facility. Response to social cues is grounded in social value orientation, the weight given by an individual to their own welfare, and the welfare of their interaction partner, in evaluation of the desirability of a particular outcome. Predicting behavior is challenging when the same social cue can be interpreted as an incentive or a deterrent. Individuals who assign positive weights to partners' welfare are classified as pro-social. Those who assign negative value or do not assign any value to their partners' welfare are characterized as pro-selves. Where pro-social individuals can interpret a behavior as a reflection of good intentions, pro-self individuals value only their own welfare and may interpret the same behavior from an interaction partner differently, as a sign of weakness. \cite{yamagishi2013behavioral} Variability in internal logical and activities that motivate farmers in particular makes a "one-size-fits-all" strategy impossible. \cite{kristensen2011} Previous work has shown the potential for social cues to contribute to behavioral flexibility, influencing the way individuals react to their environments. Performance based imitation, copying actions of those who are seen to do well, is an observed phenomenon known as social sampling. \cite{offerman2009imitation} Social cues within a computer simulated maze experiment were seen to dramatically increase the adoption of novel behavior patterns. \cite{toelch2011social} Research describing changes in operational level biosecurity compliance behavior as a result of social cues is not currently available. Identifying that social cues have an effect on operational biosecurity compliance could aid decision makers at the tactical and strategic level as they seek to increase compliance with protocols. 

We developed a novel experimental variable to treat, based on the framework of the serious game designed by Merrill et al. \cite{Merrill2019FVS}, in which participants were confronted with a compliance decision related to a common biosecurity practice, usually referred to as showering in-and-out. The shower-in, shower-out biosecurity practice, a component of the line of separation biosecurity protocol, involves changing clothes and showering before entering or exiting areas with livestock to reduce transmission of disease between animals within a facility and the outside environment. The practice is known to be highly effective at reducing the risk of infection. \cite{AndresVictorM.2015BMtC } However, workers may neglect, avoid, or insufficiently complete the practice due to the time it takes to use it. \cite{millman2017}. For example, effectively showering multiple times a day is time consuming and may be perceived to have negative repercussions such as damaging one's hair and skin or inhibiting completion of daily tasks. We developed novel experimental scenarios that tested participants' willingness to comply with the shower biosecurity practice. Specifically, we placed individuals into a farming situation where they were asked to make a binary biosecurity decision. By using the shower practice, participants incurred an associated time cost, but they avoided the potential risk of an infection. Here we hypothesize (H1) that when provided a social cue, in this case a biosecurity behavior of a coworker, individuals may react by mimicking or doing the opposite of the cue; but regardless are more likely to change their behavior in response to the cue, than if they did not receive the cue. 

Experimental variables manipulated the following factors: (1) the risk of acquiring an infection, (2) the delivery method of the infection risk, either a numerical (Numeric) value or a graphical threat gauge style image (Graphical), and (3) behavior of an automated coworker in the facility, either demonstrating compliance (Compliance by Coworker), demonstrating non-compliance (Non-Compliance by Coworker), or not demonstrating (Coworker Control). In addition to the three primary drivers of behavior, infection risk, message delivery method, and social cue; we also looked for an additional secondary driver of behavior, temporally-based psychological distancing. While we expect that participants will increase compliance in response to an infection event, this tendency to comply is likely to decrease as time passes. \cite{KahnemanDaniel1979PTAA, CarusoEugeneM2008AWiT, YiRichard2006DoPO} Referred to as a psychological distance effect, Merrill et al. \cite{ Merrill2019FVS} found that compliance did increase directly after an infection, but the effect decreased with time after the infection. We hypothesized (H2) that compliance would increase directly after an infection event, and temporally-based psychological distancing would occur defined by a decrease in compliance with increasing time since an infection. 

\section{Methods}

\subsection{Deployment}

We conducted a single experiment to examine human behavioral responses to social cues in a simulated production facility. Participants were recruited using the online workplace Amazon Mechanical Turk, which has been identified as a representative sample for the U.S. population \cite{paolacci2010running}, and a viable alternative to traditional data collection \cite{buhrmester2011amazon}. Recruits were informed that their pay would be based on performance during the experiment. Before the experiment commenced, an informational slideshow was displayed explaining the purpose of the study and mechanics of the game. This was followed by a screen allowing the recruit to choose between proceeding to play the game or declining to participate. Institutional Review Board approved practices were followed for an experiment using human participants (University of Vermont IRB \# CHRBSS-16-232-IRB).

\subsection{Experiment Design}

The simulated pork production facility was built using the Unity Development Platform (Unity Technologies, Version 5.6.3) and hosted online using WebGL \cite{parisi2012webgl}. Each participant acted as a worker and was provided information in the form of treatments that differed by combination of experimental variables: the risk of infection if they chose not to comply, the delivery method of the infection risk message, and behavior of a coworker present with them in the simulated facility. With the provided treatment information, participants were confronted with a choice to use the shower biosecurity practice or bypass the practice to avoid costs associated with usage of the practice. Each round lasting up to 70 seconds(s) represented one work day, 9am to 6pm. The experiment began with one practice round, followed by 18 rounds of incentivized play. 

\begin{figure}[H]
\centering
\includegraphics[width=25em]{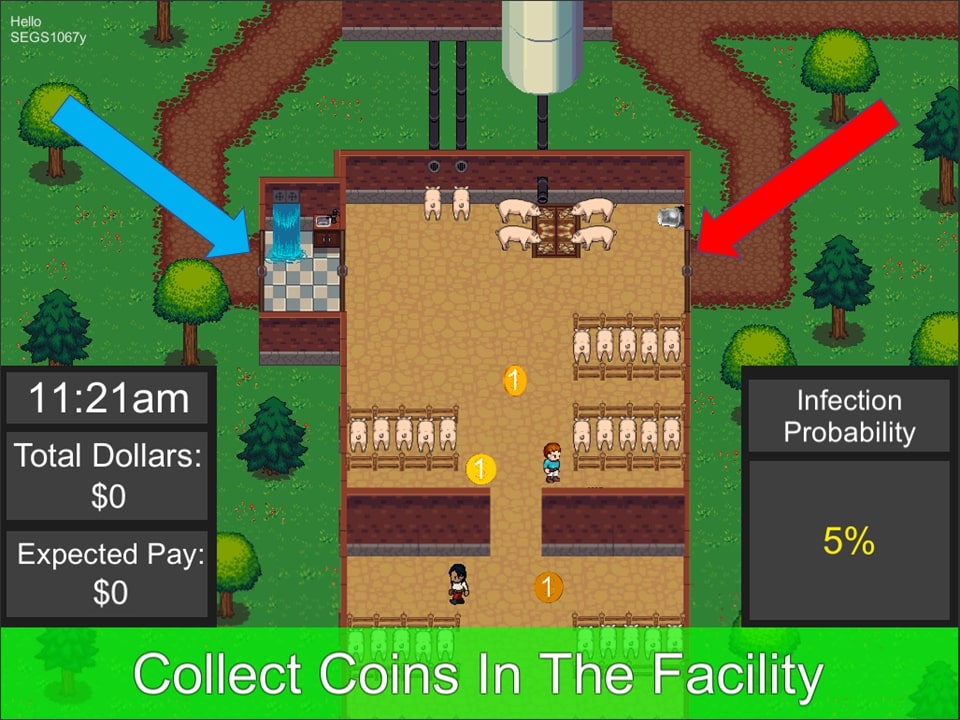}
\caption{Screenshot of a game round showing infection risk delivered as a numerical (Numeric) message, the participant-controlled worker, automated coworker, coins (internal tasks), the shower biosecurity practice (blue arrow) and emergency exit (red arrow).}
\end{figure}

To act as a worker in the facility, the participant used the computer keyboard. Each round, the worker began inside the facility. Tasks, represented as spinning coins, appeared every 2 s. When the participant moved their worker to a coin, they earned \$1 experimental dollar. Once during each round, a high-value task would appear outside the virtual facility. The value for attending to this outside task was based on the time it took to accomplish, starting at \$30 experimental dollars and decreasing by \$1 experimental dollar per second. To earn the experimental dollars for completing this high-value task, participants chose to comply and use the shower biosecurity practice which required extra time both exiting and entering the facility (with an approximate observed cost of \$8.67 experimental dollars), or avoid compliance by using the emergency exit, which carried no associated time costs but risked infection of the facility's swine. 

More specifically, across all treatments participants were asked to leave the facility to complete an outside task and were confronted with the decision of how to leave the facility, either by complying with the biosecurity practice or by skipping the practice and leaving through the emergency exit. If participants decided to use the shower biosecurity practice and "comply", participants activated a five-second counter that simulated the time it takes to shower and change clothes. After 5 s the virtual worker could exit the shower and complete the outside task. The same procedure, with another 5 s delay, was repeated upon re-entry of the facility post-task completion. If the participant decided on non-compliance with the shower biosecurity practice, and thus, left the facility through the emergency exit, they incurred no time cost, but there was an associated chance of infection based on the actual infection risk probability during the given round. The risk of using the emergency exit was quantified using the infection information presented to the participant, which varied by treatment (see Table 1 for breakdown of information provided to participants by experimental variable and associated treatment levels). If an infection occurred, calculated using a pseudorandom number generator, the round ended immediately, and the participant lost \$50 experimental dollars as well as any expected payout they had collected during the round. If an infection did not occur, the round continued until the normal end of the workday. The mean observed opportunity cost of using the shower practice as opposed to the emergency exit was calculated to be \$8.67 experimental dollars, due to the time lost that could have been used to complete the outside task more quickly and collect coins inside the facility upon reentry. 

In addition to the participant's worker, an automated coworker was included in all simulation scenarios to provide implicit social cues to the player. The only explicit information given to the participants regarding the coworker was during the pre-game slide show (see Supplementary Materials) where they read that there may be another worker in the facility with them. The automated coworker was completely predetermined in its actions, exhibiting one of three behaviors in any given round: 1) compliance with the shower biosecurity practice, 2) use of the emergency exit, or 3) not exiting the facility (Control). During the coworker compliance and non-compliance treatments, participants would see the decision made by the coworker, before they made their own decision to use the shower biosecurity practice or the emergency exit. This demonstration by the coworker, an implicit social cue, was intended to be observed by the participant with the potential to affect their decision to comply. 

After each round, the number of experimental dollars earned within the round was displayed on the participant's screen. In addition, a cumulative sum of the total experimental dollars earned thus far in the experiment was displayed. At the end of the 18 experimental rounds, participants received \$1 U.S. for each \$350 experimental dollars plus a base pay of \$3.00 U.S.

\subsection{Experimental Variables and Treatments}

Experimental variables were designed to test factors shown to influence human behavior: risk of infection, message delivery method, and social cue. Participants were confronted with infection risk, delivered in different message formats, at the start of each round. This infection risk information was used by the participant during the round when they chose to either comply with the shower biosecurity protocol, or risk using the emergency exit for a slightly higher payout. The infection risk information was delivered numerically (e.g., "5\% infection risk"), or graphically using a threat gauge (Figure 2). 

\begin{figure}[H]
\begin{adjustwidth}{-2cm}{}
\subfloat[][]{\includegraphics[width=23em]{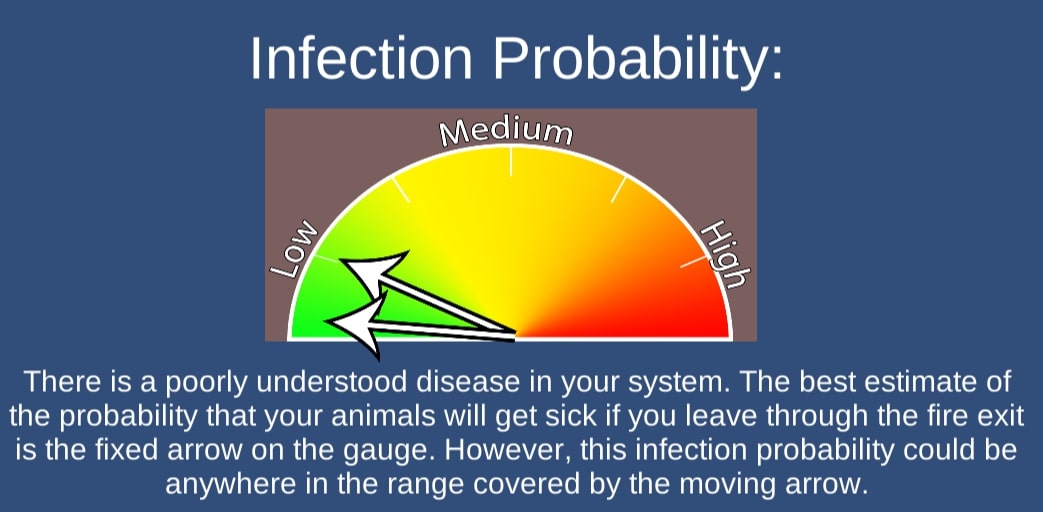}}
\vspace{1cm}
\subfloat[][]{\includegraphics[width=23em]{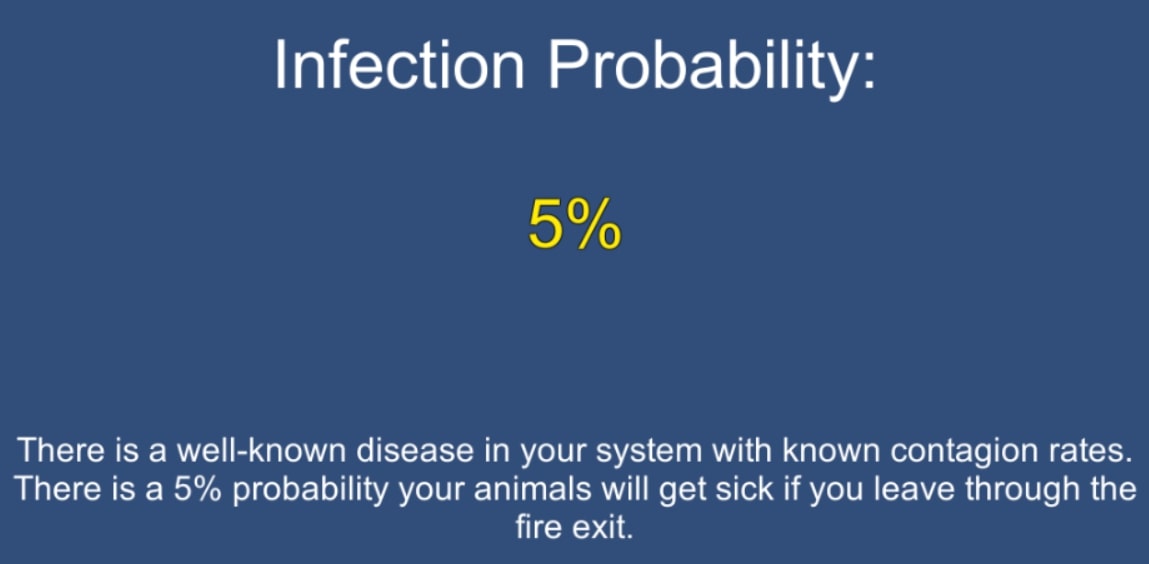}}
\end{adjustwidth}
\caption{Start of round infection risk delivered as a graphical (Graphical) message; two arrows, one fixed, the other moving, were used to convey a best estimate and uncertainty around that estimate (a). Start of round infection risk delivered as a numerical (Numeric) message (b).}
\end{figure}

In the original work of Merrill et al. \cite{Merrill2019FVS}, the certainty of the disease risk information and the message delivery method were both treated as separate variables. When disease infection risk was treated with certainty, participants were provided a single infection probability. Disease infection risk was treated with uncertainty by providing a best estimate of the infection probability in addition to a range of potential values. It was determined that numerical values \textit{with certainty} were the most likely to be associated with avoidance of the biosecurity practice, and graphical threat gauge style images \textit{with uncertainty} were the best at increasing willingness to comply. \cite{Merrill2019FVS} In order to increase our sample size per treatment, we utilized only these two types of message delivery methods: numeric with certainty, and graphical with uncertainty, respectively. Hereafter, these are referred to as Numeric and Graphical. 

For our novel social cue experimental variable, an automated coworker demonstrated one of three behaviors prior to the participant's decision to comply. Unbeknownst to the participant, the compliance decision chosen by the coworker was predetermined by treatment. Coworker demonstration of compliance involved the automated coworker using the shower biosecurity practice; likewise, the coworker demonstration of non-compliance involved the coworker using the emergency exit. No demonstration, the control treatment, indicates that the coworker never left the facility. 

In summary, the experiment had three infection risk treatments: (very low (1\%), to low (5\%), to medium (15\%)), two infection risk messaging treatments (Numeric and Graphical), and three social cue treatments (Compliance by Coworker, Non-Compliance by Coworker, and Coworker Control). A complete block design was utilized, in which data were collected for every combination of the three experimental variables and their associated levels. Two additional variables were also used in this experiment. First, the psychological distance effect, to identify changes in behavior related to experiencing infection events. The distancing effect was quantified by simply counting the number of rounds since an infection event occurred. Second, because the experiment takes place in a series of rounds, we used a variable referred to as play order to control for within-experiment learning. \cite{LevinthalDa1993TMOL}

\begin{table}[H]
\centering
\begin{tabular}{|l|l|}
\hline
\multicolumn{1}{|l|}{\textbf{Treatment}} & \multicolumn{1}{l|}{\textit{\textbf{N}}} \\ \hline
Infection Risk: 1\% (Very Low) & 648 (6 rounds * 108 participants) \\ \hline
Infection Risk: 5\% (Low) & 648 (6 rounds * 108 participants) \\ \hline
Infection Risk: 15\% (Medium) & 648 (6 rounds * 108 participants) \\ \hline
\begin{tabular}[c]{@{}l@{}}Message Delivery Method: \\ Numeric ("1\%," "5\%," or "15\%") \end{tabular} & 972 (9 rounds * 108 participants) \\ \hline
\begin{tabular}[c]{@{}l@{}}Message Delivery Method: \\ Graphical (A threat gauge with \\ arrows used to indicate risk)
\end{tabular} & 972 (9 rounds * 108 participants) \\ \hline
\begin{tabular}[c]{@{}l@{}} Social Cue: \\ Compliance by Coworker
\end{tabular} & 648 (6 rounds * 108 participants) \\ \hline
\begin{tabular}[c]{@{}l@{}} Social Cue: \\ Non-Compliance by Coworker
\end{tabular} & 648 (6 rounds * 108 participants) \\ \hline
\begin{tabular}[c]{@{}l@{}} Social Cue: \\ Coworker Control
\end{tabular} & 648 (6 rounds * 108 participants) \\ \hline
\end{tabular}
\caption{Experiment treatments.}
\end{table}

\subsection{Analysis}

\subsubsection{Linear effects}

The response variable in our experiment is binary, either the participant used the shower biosecurity practice, or avoided compliance by using the emergency exit. To explain the response variable, a set of mixed-effect logistic regression candidate models were generated using the statistical programming language, R. \cite{ r-core2019, wickham2016, bates2014fitting} The set of candidate models included mixtures of experimental variables (Table 1), interaction terms between experimental variables, as well as predictor variables: (1) psychological distance, and (2) play order. Participant was added as a random effect in all models to account for variation between individuals. To identify the model of best fit, we used an information theoretic approach to test how well each of our candidate models explains the data. \cite{BurnhamKennethP2002Msam, burnham2004multimodel} Models were evaluated using Akaike's Information Criterion (AIC), where the lowest AIC value indicates the most parsimonious candidate model that best explains variation in the response variable with the fewest parameterized variables. \cite{akaike1973information} 

\subsubsection{Ratio of variances}

Nonlinear effects resulting from inconsistent responses to the social cue experimental variable are identified in a separate examination of the strategy variance components. Here we quantify strategic variability by calculating how an individual changed their compliance strategy between the three social cue treatments: compliance, non-compliance, and control. Due to our complete block experiment design, we can isolate an individual's average compliance in response to any one of our three social cue treatments by averaging over the other two experimental variables: infection risk and message delivery method. Having identified an individual's average compliance strategy for each of the three social cue treatments they were presented, we then calculate the differences in strategy between each of the three pairs of social cue treatments (compliance vs. non-compliance, compliance vs. control, non-compliance vs. control). For example, how did an individual's average compliance change between all rounds in which they were presented the social cue compliance treatment, versus all rounds they were presented with the social cue non-compliance treatment. 

After quantifying individuals' strategic variability using changes in average compliance, we can aggregate all participants' average compliance changes into distributions, grouped by the pairs of social cue treatments between which individuals' may have collectively changed their average behavior. The variance of each of the three distributions can be evaluated in relation to one-another as ratios of variances, which quantify differences in strategic variability observed over all individuals. Confidence intervals were calculated using an F-test to determine if any of the ratio of variances are statistically significant.

\section{Results}

Data were collected from 108 participants for the experiment, the average payout was \$5.62 U.S. with a minimum of \$4.32 U.S. and a maximum of \$6.45 U.S. From the set of mixed-effect logistic regression candidate models, the model with the lowest AIC, Model 1, was selected as the best supported candidate model and used for reported statistical inference from the experiment (Table 2). Model 1 included the fixed effects psychological distance, message delivery method, infection risk, and play order, in addition to participant as a random effect. $\Delta$AIC quantifies the loss of information if a different candidate model is used. Neither the social cue experimental variable nor interaction effects were included in the AIC-selected best candidate model. Interaction effects between experimental variables were not found to be significant in any of the candidate models generated. 

\begin{table}[H]
\centering
\begin{tabular}{|l|c|c|c|c|c|c|c|c|r|r|}
\hline
\textbf{Model} & \textbf{PD} & \textbf{M} & \textbf{IR} & \textbf{PO} & \textbf{SC} & \textbf{M*IR} & \textbf{M*SC} & \textbf{SC*IR} & \textbf{AIC} & \textbf{$\Delta$AIC} \\ \hline
1 & \textbf{X} & \textbf{X} & \textbf{X} & \textbf{X} & \textbf{} & \textbf{} & \textbf{} & \textbf{} & 1495 & 0 \\ \hline
3 & \textbf{X} & \textbf{X} & \textbf{X} & \textbf{X} & \textbf{} & \textbf{X} & \textbf{} & \textbf{} & 1495 & 0.280 \\ \hline
2 & \textbf{X} & \textbf{X} & \textbf{X} & \textbf{X} & \textbf{X} & \textbf{} & \textbf{} & \textbf{} & 1498 & 3.081 \\ \hline
7 & \textbf{X} & \textbf{X} & \textbf{X} & \textbf{X} & \textbf{X} & \textbf{X} & \textbf{} & \textbf{} & 1499 & 3.383 \\ \hline
4 & \textbf{X} & \textbf{X} & \textbf{X} & \textbf{X} & \textbf{} & \textbf{X} & \textbf{X} & \textbf{} & 1502 & 6.764 \\ \hline
5 & \textbf{X} & \textbf{X} & \textbf{X} & \textbf{X} & \textbf{} & \textbf{} & \textbf{} & \textbf{X} & 1505 & 10.195 \\ \hline
6 & \textbf{X} & \textbf{X} & \textbf{X} & \textbf{X} & \textbf{} & \textbf{X} & \textbf{} & \textbf{X} & 1506 & 10.523 \\ \hline
\end{tabular}
\caption{Candidate Models reordered by AIC value with the best AIC-selected models listed first. Independent variables: Psychological Distance (PD), Message Delivery Method (M), Infection Risk (IR), Play Order (PO) and Social Cue (SC). Interaction terms e.g. Message Delivery Method by Infection Risk are denoted as (M*IR).}
\end{table}

The models use a linear combination of the random and fixed effects to obtain logit coefficients that predict the response variable, or the probability (from 0 to 1), that the participant would comply with the shower biosecurity practice. We exponentiated the logit coefficients to generate odds ratios, which were used to evaluate the odds that an individual will opt to use the shower biosecurity practice as opposed to the emergency exit. An odds ratio of 1:1, presented as 1, indicated that there are even odds for either outcome. If an odds ratio confidence interval excluded one, the variable was considered significant. Odds ratios that were greater than 1 indicated that it was significantly more likely that the participant complied with the shower practice than used the emergency exit; ratios below 1 indicated it was more likely the participant used the emergency exit. 

Results from the logistic regression quantify the model-predicted probability that the participant will comply with the shower biosecurity practice. Presented as odds ratios in Table 3, the first row represents the baseline odds ratio (intercept) associated with the treatment combination of 5\% infection risk delivered using a Graphical message. The 66.536 odds ratio (intercept) signifies participants are 66.536 times as likely to use the shower biosecurity practice as opposed to the emergency exit when provided with the Graphical message at the 5\% infection risk level. The rest of the odds ratios in Table 3 are compared to the baseline, intercept ratio. For example, participants that received the infection risk information as a Numeric message as opposed to the intercept message delivery method (Graphical), had an odds ratio of 0.095. Therefore, participants receiving a Numeric message were 0.095 times as likely to use the shower biosecurity practice, or 10.526 times as likely to use the emergency exit, than participants receiving a Graphical message. No conclusion can be drawn about how predictor variables affect compliance decisions if the odds ratio confidence interval includes 1. We found significant main effects (Table 3, Figure 3). 

\begin{table}[H]
\centering
\begin{tabular}{|l|c|c|c|r|}
\hline
\textbf{Parameter} & \textbf{Odds ratio} & \textbf{Lower bound} & \textbf{Upper bound} & \textbf{\textit{P}-value} \\ \hline
\begin{tabular}[c]{@{}l@{}}Intercept \\(Graphical Message, \\ Infection Risk @ 5\%)\end{tabular} 
& 66.536 & 22.729 & 194.774 & \textbf{\textless{}0.001} \\ \hline
Psychological Distance & 0.120 & 0.048 & 0.299 & \textbf{\textless{}0.001} \\ \hline
Numeric Message & 0.095 & 0.068 & 0.134 & \textbf{\textless{}0.001} \\ \hline
Infection Risk @ 15\% & 20.100 & 12.885 & 31.353 & \textbf{\textless{}0.001} \\ \hline
Infection Risk @ 1\% & 0.116 & 0.081 & 0.165 & \textbf{\textless{}0.001} \\ \hline
Play Order & 0.976 & 0.949 & 1.004 & 0.090 \\ \hline
\end{tabular}
\caption{Results of the selected best fit mixed-effect logistic regression model (Model 1; see Table 2). Depicted here are the odds ratios for fixed effects describing relationships with the binary response variable: compliance with the biosecurity practice. Bold values indicate significance at $\alpha$ = 0.05.}
\end{table}

\subsection{Main Effects}

Average compliance increased with increasing infection risk from 1\% (34\% compliance) to 5\% (60\% compliance) to 15\% (89\% compliance) (Figure 3 Left Panel). Compliance was also much higher when the infection risk message was delivered as a Graphical message (73\% compliance) vs. as a Numeric message (49\% compliance) (Figure 3 Right Panel).

\begin{figure}[H]
\centering
\includegraphics[width=32em]{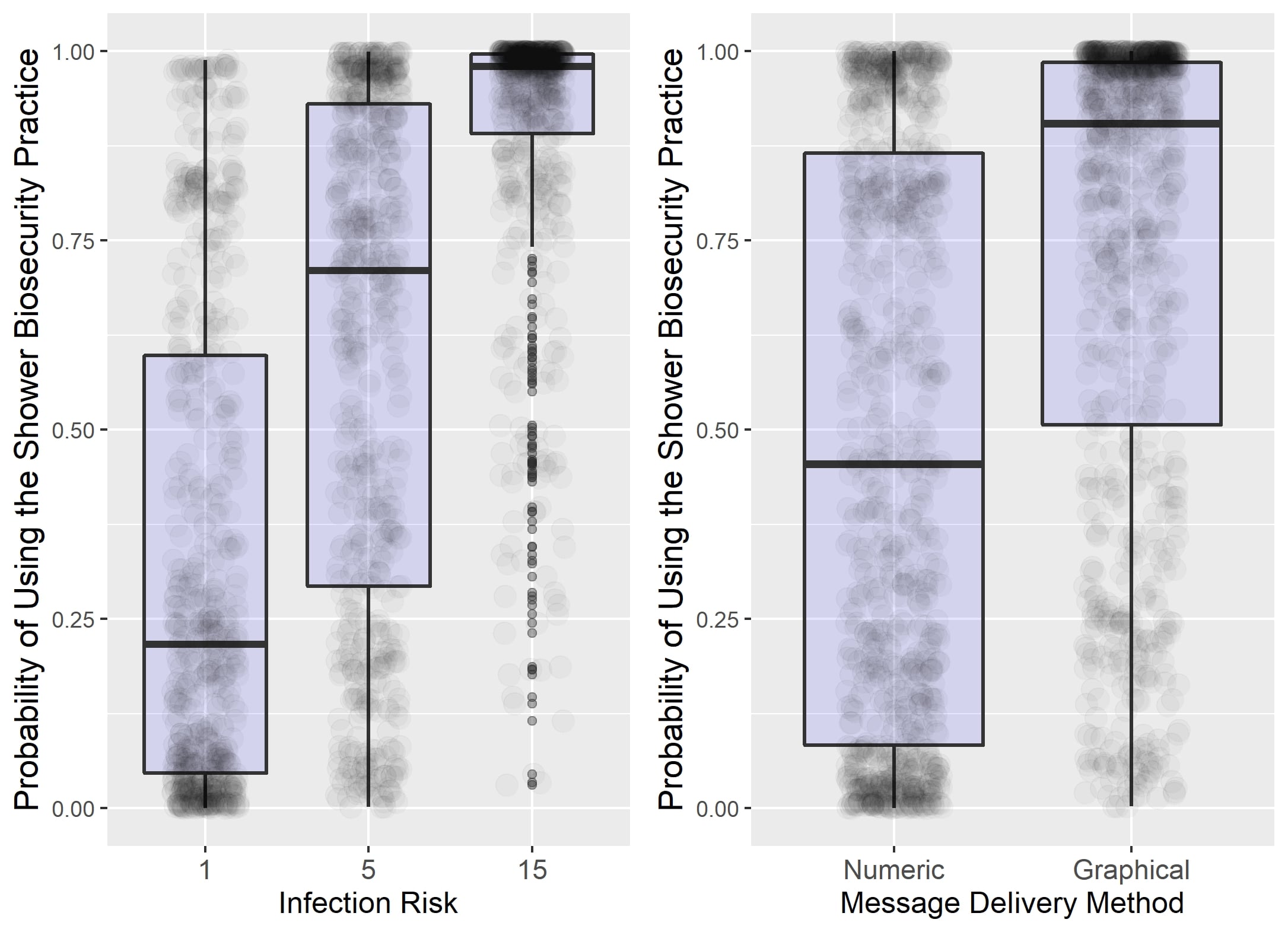}
\caption{Summary results of the main treatment effects. Box-plot of the probability of using the shower biosecurity practice by the main effects, Message Delivery Method and Infection Risk. Lower and upper box boundaries 25th and 75th percentiles, respectively, line inside box median, overlaid on model predicted data values.}
\end{figure}

\subsection{Individual Differences (H1) }

Strategic variability of individual behavior was quantified to determine if the coworker social cues elicited any measurable response. The social cue variable was not included in the AIC-selected best candidate model (Table 2). This indicated that individuals did not respond in a consistent or linear way to the coworker demonstrations, so the information gained by including the social cue variable did not explain enough information to overcome the penalty for inclusion of an additional parameterized variable. While the changes in compliance were not consistent across participants, we sought to discover if the coworker behavior influenced the strategies used by the participants. We identified the strategic variability of individuals by quantifying their change in average compliance when confronted by the different coworker behaviors. Controlling for the effects of the infection risk and message delivery method experimental variables, we calculate individual changes in average compliance between the three social cue treatments: Compliance by Coworker, Non-Compliance by Coworker, and Coworker Control. Specifically, we identify the distributions of average changes in compliance between the three \textit{combinations} of these treatments: Compliance by Coworker vs. Non-Compliance by Coworker, Compliance by Coworker vs. Coworker Control, and Non-Compliance by Coworker vs. Coworker Control (Figure 4). This allowed us to test if the strategy space was distributed differently between the social cue treatments.

\begin{figure}[H]
\centering
\includegraphics[width=35em]{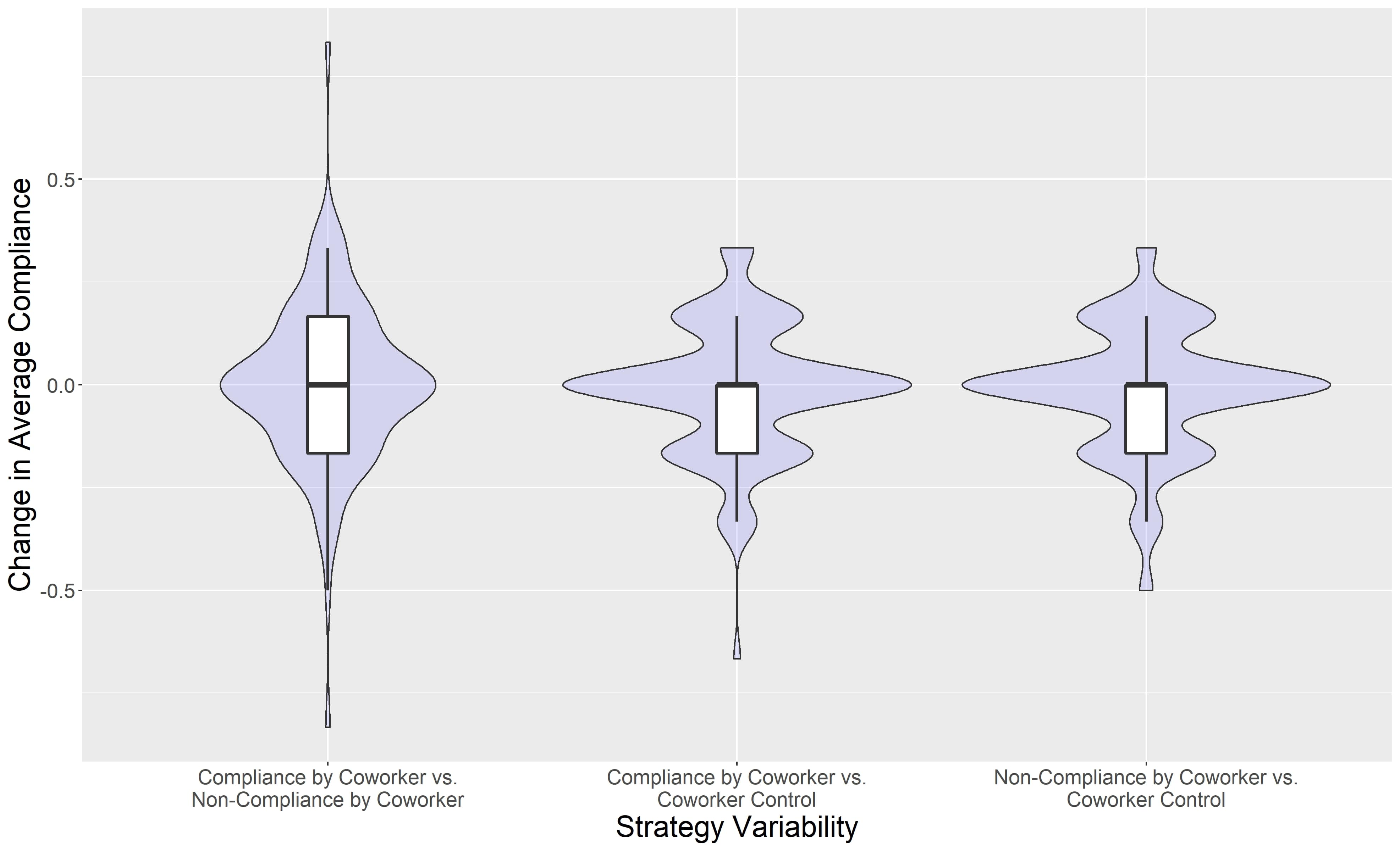}
\caption{ Violin with inlaid box-plots of individuals' changes in average compliance between coworker treatments. Lower and upper box boundaries 25th and 75th percentiles, respectively, line inside box median.}
\end{figure}

Change in average compliance, normalized to span from -1 to 1, represented the degree to which an individual changed their compliance strategy between two types of social cue treatments. Of particular interest was whether participants altered their strategies when observing a coworker exiting the production facility (either following protocols or not) as contrasted with the strategies they employed when the coworker never left the facility. The greatest strategic variability was observed in the changes of compliance between Compliance by Coworker treatments and Non-Compliance by Coworker treatments (Figure 4, Left-most violin with inlaid box-plot). In this case, participants with a positive change in compliance (e.g. increasing compliance from a base rate), falling above the y=0 line, were complying with the shower biosecurity practice more often during Compliance by Coworker treatments, when compared with Non-Compliance by Coworker treatments. However, there were also individuals doing the opposite; participants falling below the $y=0$ line were complying more often during Non-Compliance by Coworker treatments, when compared with Compliance by Coworker treatments. To evaluate the relative change in variance between the three distributions of compliance strategies (Figure 4), we enumerated their ratio of variances (Table 4). 

\begin{table}[H]
\begin{adjustwidth}{-2cm}{}
\begin{tabular}{ll|c|c|l}
\cline{3-4}
\multicolumn{1}{c}{\textbf{}} & \textbf{} & \multicolumn{2}{c|}{\textbf{Ratio of variances}} & \textbf{} \\ \cline{1-4}
\multicolumn{1}{|r|}{\makecell{Compliance by Coworker vs.\\Non-Compliance by Coworker }} & \multicolumn{1}{c|}{0.040} & 1.496 & 1.680 & \\ \cline{1-4}
\multicolumn{1}{|r|}{\makecell{ Compliance by Coworker\\vs. Coworker Control}} & \multicolumn{1}{c|}{0.027} & 1 & 1.123 & \\ \cline{1-4}
\multicolumn{1}{l|}{} & \textbf{Variance} & 0.027 & 0.024 & \\ \cline{2-4}
& & \multicolumn{1}{l|}{\makecell{ Compliance by Coworker \\vs. Coworker Control}} & \multicolumn{1}{l|}{\makecell{ Non-Compliance by Coworker \\vs. Coworker Control}} & \\ \cline{3-4}
\end{tabular}
\end{adjustwidth}
\caption{Variance of changes in average compliance between the three combinations of social normative behavior treatments. Ratios of variances quantify the relative change in variance between two distributions.}
\end{table}

To determine if any of the ratio of variances were statistically significant, confidence intervals were calculated using an F-test (Table 5). A ratio was determined to be significant if the confidence interval did not include 1. The top two rows of Table 5 with significant $p$-values correspond to the increased variation in individual compliance strategies observed between the two explicit demonstrations (Compliance by Coworker vs. Non-Compliance by Coworker); with respect to compliance strategy variation between either of the two explicit demonstrations and the baseline treatment (Compliance by Coworker vs. Coworker Control, or Non-Compliance by Coworker vs. Coworker Control). This confirmed the hypothesis (H1) that participants would change their behavior using the social cues presented to them. In this case the change in behavior came in the form of increased strategic variability between either of the two explicit social cues with respect to the control. The third row indicates there was not a significant difference in strategic variability between Compliance by Coworker vs. Coworker Control, and Non-Compliance by Coworker vs. Coworker Control. 

\begin{table}[H]
\centering
\begin{tabular}{|c|c|c|c|r|}
\hline
\makecell{\textbf{Paired Social Cue}\\\textbf{Treatment Ratio}}&\makecell{\textbf{Ratio of}\\\textbf{ variances}} &\makecell{ \textbf{Lower}\\\textbf{ bound}} & \makecell{\textbf{Upper}\\\textbf{ bound}} & \textbf{\textit{P}-value} \\ \hline
(SC1 vs. SC2) / (SC1 vs. Control) & 1.496 & 1.022 & 2.190 & \textbf{0.038} \\ \hline
(SC1 vs. SC2) / (SC2 vs. Control) & 1.680 & 1.148 & 2.459 & \textbf{0.008} \\ \hline
(SC1 vs. Control) / (SC2 vs. Control) & 1.123 & 0.767 & 1.643 & 0.550 \\ \hline
\end{tabular}
\caption{ Results of F-test to calculate ratio of variances (Table 4) confidence intervals. SC1 corresponds with Compliance by Coworker treatments, SC2 corresponds with Non-Compliance by Coworker treatments, Control corresponds with Coworker Control treatments. Bold indicates statistical significance, $\alpha$ = 0.05.}
\end{table}

\subsection{Psychological Distance (H3) }

The results of the AIC-selected best fitting model from the experiment found the psychological distance effect to be significant with an odds ratio of 0.120 (Table 3). This confirms (H2) that temporally-based psychological distancing occurred by indicating that the probability that individuals complied with the shower biosecurity practice increased directly after becoming infected, and the effect decayed with time. 

\section{Discussion}

The goal of our experiment was to investigate the effects of social cues in a simulated pork production facility. Our study advances knowledge of how heuristics \cite{TverskyA1974JuUH}, or affective reasoning \cite{SlovicPaul2004RaAa, SlovicPaul2005ARaD}, may impact decisions made under time pressure. We examined the effects of introducing social cues by testing for disease infection risk, message delivery method, and social cues. Our results unveil the potential impacts of social value orientation \cite{yamagishi2013behavioral} by identifying effects of different social cues at the operational level of biosecurity. Social cues were shown to have a significant effect on the degree of strategic variability in compliance behavior (H1). That is, people reacted and changed their behavior when they observed the coworker's biosecurity decision as compared to when the coworker staying in the facility for the entire working day. We also found evidence of a psychological distance effect (H2), an increased likelihood to comply directly after an infection event, an effect that decayed over time. 

\subsection{Limitations}

Statistical sampling methods are often used to study selected groups and strata of society, but they do carry the potential to introduce bias into an experiment when the sampled population has marked differences from the target population. For our study, we assume that results obtained using the online platform Amazon Mechanical Turk can be extrapolated to the target population of swine industry professionals. Amazon Mechanical Turk has been identified as a viable alternative to traditional sampling methods like surveys \cite{buhrmester2011amazon}, and is also characterized as a more representative sample of the U.S. national population than either college undergraduate or internet samples in general \cite{paolacci2010running}. Although limited research exists to compare the differences between swine industry workers and the general population, an experiment by Clark et al. \cite{clark2019using} did not detect a difference in the distributions of risk behavioral strategies between a sample of online participants and agricultural professionals. Farmers in particular exhibit variable responses to stimuli due to the complexity of the decision-making process \cite{kristensen2011}; and because they are operating under a variety of different objective functions a consistent bias is unlikely. 

Merrill et al. \cite{Merrill2019FVS}, specifically their second experiment, was the foundation for the experiment within the current study. It is important to acknowledge the differences in the current experiment, both to enhance replicability, as well as to address any potential bias in comparison. In the current experiment with respect to the work of Merrill et al. \cite{Merrill2019FVS}: the linguistic message delivery treatment was removed, the social cue experimental variable was introduced, the appearance of the threat gauge was adjusted, the experiment had six fewer rounds, and each round was approximately 10 seconds longer. Although the look of the threat gauge was modified, the relatively consistent use of verbal demarcations (Low, Medium, High), should preserve response to the graphical message across experiments. 

\subsection{Infection Risk}

Infection risk has been determined to be the main driver of behavior both in this current experiment, as well as the work of Merrill et al. \cite{ Merrill2019FVS}. In the study by Merrill et al. \cite{ Merrill2019FVS}, participants were exposed to infection risk treatments of very low (1\%), low (5\%), medium (15\%) and high (25\%). In the current experiment, the high infection risk treatment with a 25\% chance of infection was omitted because Merrill et al. \cite{ Merrill2019FVS} observed near ubiquitous compliance at 25\%, and we wanted to increase our ability to detect differences in other signals, such as the social cue experimental variable. Therefore, participants were only confronted with the very low (1\%), low (5\%), and medium (15\%) infection risk treatments. The change in risk perception that could be attributed to the removal of the high infection risk treatment can be seen when comparing average compliance in the current experiment, with average compliance in analogous treatments from the experiment of Merrill et al. \cite{ Merrill2019FVS} (Table 6). Here analogous refers to message delivery method treatments in the experiment conducted by Merrill et al. \cite{ Merrill2019FVS} that were replicated in the current experiment, numeric with certainty and graphical with uncertainty; with message delivery method treatments that were not replicated, e.g. numeric with uncertainty, omitted from comparison. 

\begin{table}[H]
\centering
\begin{tabular}{|c|c|c|c|}
\hline
\makecell{\textbf{Infection}\\\textbf{risk}} & \begin{tabular}[c]{@{}c@{}}\makecell{\textbf{Merrill et al.}\\\textbf{frequency of}\\\textbf{compliance (\%)}}\end{tabular} & \begin{tabular}[c]{@{}c@{}}\makecell{\textbf{Current study}\\\textbf{frequency of}\\\textbf{compliance (\%)}}\end{tabular} & \makecell{\textbf{$\Delta$ frequency of}\\\textbf{compliance (\%)}} \\ \hline
1 & 23.6 & 33.6 & 10.0 \\ \hline
5 & 56.0 & 60.3 & 4.3 \\ \hline
15 & 80.9 & 88.7 & 7.8 \\ \hhline{|=|=|=|=|}
\textbf{Average} & 53.5 & 60.9 & 7.4 \\ \hline
\end{tabular}
\caption{Change in observed frequency of use of the shower biosecurity practice between original analogous treatments and current experiment treatments by infection risk. $\Delta$ frequency of compliance quantifies the change between the original and current average compliance, positive values correspond to an increase in the current study average compliance. }
\end{table}

At analogous treatments in the experiment of Merrill et al. \cite{ Merrill2019FVS}, average compliance increased from 23.6\% at very low infection risk, to 56.0\% at low risk, to 80.9\% at medium risk, to 91.0\% at high risk. We found increased compliance with the shower biosecurity practice in our experiment. In the current study, average compliance increased from 33.6\% at very low infection risk, to 60.3\% at low infection risk, to 88.7\% at medium infection risk. By removing the high infection risk treatment, participants may have exhibited increased risk aversion in decision-making because they were now using the medium infection risk treatment as their point of reference for maximum infection risk. This effect of reframing infection risk is consistent with heuristics identified in risk perception literature \cite{ TverskyA1974JuUH, KahnemanDaniel1979PTAA}. It is also possible that changes in observed compliance between experiments could have been caused by the introduction of the social cue experimental variable. We propose an additional experiment that utilizes our novel social cue experimental variable, without omitting the 25\% risk level. This potential future work could discern some of these subtleties of risk communication strategies, and determine if it was the introduction of the social cue, or the omission of the high infection risk level, that caused the observed increase in compliance. 

\subsection{Individual Differences (H2)}

By introducing the social cue variable into the experiment design, we hypothesized that compliance behavior could be shifted by an implicit suggestion. In this case, the social cue came from a coworker present in the facility with the participant. We observed increased strategic variability by participants in their response to compliance or non-compliance by the coworker, with respect to the control where the coworker never exited. The behaviors of the automated coworker could have changed how a participant perceived what is commonly done in the facility. Although the automated coworker had no effect on the actual outcome of the round, we expected the subtle hint of how the coworker handled the compliance decision to be internalized in the decision-making process of participants. Humans are known to let feelings and mental shortcuts guide the decision-making process, especially when under time pressure. \cite{SlovicPaul2004RaAa, SlovicPaul2005ARaD, TverskyA1974JuUH }

In our examination of the variability in the participants' strategy space, we established that individuals displayed significantly more variation in compliance strategy when responding to two different explicit social behaviors: Compliance by Coworker and Non-Compliance by Coworker. The social value orientation of participants, associated with varying tendencies to cooperate or compete with interaction partners, likely played a role in observed differences in strategic variability. \cite{yamagishi2013behavioral} These differences could also be attributed to a mirroring or mimicry effect; imitation based on social performance cues has been observed, although in this experiment there was no effect on performance or payout related to the cue. \cite{offerman2009imitation} This aligns with research identifying the impact of social cues on behavioral flexibility, leading to novel behavioral patterns. \cite{toelch2011social} Quantified as variance, or ratio of variances, having a significant increase indicates individuals were changing, but not in a consistent way. One explanation is that individuals are known to be motivated by a wide range of factors and internal logic. \cite{kristensen2011} The complex decision-making process in each individual may lead to varying degrees of "follower" or "anti-follower" tendencies. If a worker in a facility sees their associate breaking the rules, are they more likely to pick up the slack and "cover for them", or are they more likely to feel it is justified for them to break the rules as well? One player who shared their strategy after the experiment stated they would always use the shower biosecurity practice when the coworker was exiting the emergency door to compensate/offset their coworker's bad behavior. They felt more inclined to shower in order to protect the farm from disease. 

Identifying that social cues influence compliance strategy at the operational level is a novel conclusion and lends credence and support to future studies of worker culture. We recommend follow-up experiments with more explicit social cues, e.g., where a coworker demonstration is paired with an associated explanation of behavior communicated via a text bubble. For example, how would pairing the coworker demonstration of compliance within this experiment with an explicit verbal explanation from the coworker (e.g. "Showering takes time, but we are all in this together.") affect behavior of participants? Another avenue to further research effects of social cues on operational biosecurity compliance is to replicate the current experiment in a virtual-reality based swine production facility, in order to test if the results are consistent in a more immersive environment. Continuing to study how social cues can increase compliance, or induce non-compliance, will be valuable for farm management, training, and monitoring. 

\subsection{Psychological Distance (H3)}

In our experiment, individuals who had recently incurred a facility infection were far more likely to comply with the shower biosecurity practice. Over time, mental construal of events are expected to become more abstract and low-level, decreasing their influence on future decisions. \cite{ trope2003temporal} In the context of this experiment, the increased likelihood to comply directly after an infection event wore off over time. This temporally based psychological distancing \cite{CarusoEugeneM2008AWiT, YiRichard2006DoPO} was similarly observed by Merrill et al. \cite{Merrill2019FVS}, where individuals who had just been infected were twice as likely to comply as those who had never been infected. Our results therefore suggest that it is important to acknowledge that temporal psychological distancing will influence workers and the decrease in observed compliance should be combatted such as with biosecurity trainings that are reinforced frequently. 

\section{Conclusion}

The human behavioral component of animal biosecurity is not well understood, but worker decisions at the operational level have direct economic and sociological consequences when an outbreak occurs. \cite{ritter2017invited, RacicotManon2011Do4b } For both managers and workers in the swine industry, understanding how social cues are affecting compliance with existing biosecurity protocols is critical. This study demonstrates the ability to test hypotheses about human behavioral responses to social cues using experimental game simulations. We examined how a coworker demonstration can impact the complex mental process occurring when participants chose to comply with the simulated shower biosecurity practice in the experiment. We tested the effect of an implicit social cue on decision-making; an attempt to subtly recreate the worker culture in a facility. \cite{ AjzenIcek1991Ttop} While it is well known that workers do not always comply with operational-level protocols \cite{ RacicotManon2011Do4b }, there is no existing data that shows how social cues may be impacting these biosecurity lapses. These findings outline the significant variability in how people change their behavior in response to those around them, even when others' actions have no repercussions or effect on monetary payout.

To summarize and extend these findings to be applicable at the facility level, this work identifies that social cues will result in different compliance strategies amongst workers in a swine facility. Behavior of others, even if not directly impacting the worker, will impact their decision making. Providing this feedback from the operational level can inform tactical and strategic decision makers as they implement biosecurity protocols whose efficacy will depend on workers' willingness to comply, and attempt to create a workplace culture of compliance. Humans are extremely complex, and while we acknowledge that there is no blanket solution to increasing worker compliance, we believe small changes can have an impact on the system level. Advancing knowledge and understanding of human behavioral components of animal biosecurity has vast potential to increase worker and animal welfare, shifting the industry towards disease-resiliency. 

\bibliography{Trinity_bibliography}
\bibliographystyle{unsrt}
\end{document}